\documentclass[]{aastex631}

\graphicspath{{./}{figures/}}

\begin{document}

\title{QLP Data Release Notes 001: K2 + TESS Analysis}

\correspondingauthor{Michelle Kunimoto}
\email{mkuni@mit.edu}

\author[0000-0001-9269-8060]{Michelle Kunimoto}
\affiliation{Kavli Institute for Astrophysics and Space Research, Massachusetts Institute of Technology, Cambridge, MA 02139}

\author[0000-0003-0918-7484]{Chelsea Huang}
\affiliation{Kavli Institute for Astrophysics and Space Research, Massachusetts Institute of Technology, Cambridge, MA 02139}

\author[0000-0002-5308-8603]{Evan Tey}
\affiliation{Kavli Institute for Astrophysics and Space Research, Massachusetts Institute of Technology, Cambridge, MA 02139}

\author[0000-0003-0241-2757]{Willie Fong}
\affiliation{Kavli Institute for Astrophysics and Space Research, Massachusetts Institute of Technology, Cambridge, MA 02139}

\author[0000-0002-2135-9018]{Katharine Hesse}
\affiliation{Kavli Institute for Astrophysics and Space Research, Massachusetts Institute of Technology, Cambridge, MA 02139}

\author[0000-0002-1836-3120]{Avi Shporer}
\affiliation{Kavli Institute for Astrophysics and Space Research, Massachusetts Institute of Technology, Cambridge, MA 02139}

\begin{abstract}
TESS will target the ecliptic plane in Sectors 42 -- 46. These sectors overlap with campaigns from the K2 mission, providing a unique opportunity for multi-mission light curve analysis. This data release note describes the combined analysis of K2 and TESS light curves as part of the Quick-Look Pipeline (QLP) procedure, which processes light curves for all targets in TESS Full-Frame Images (FFIs) down to TESS magnitude $T = 13.5$. We describe updates to our codebase, and the planet transit search, candidate triage, and report generation that are affected by this combined analysis.
\end{abstract}

\keywords{Exoplanets (498) --- Exoplanet detection methods (489) --- Transit photometry (1709) --- Time series analysis (1916)}

\section{Updates}

\subsection{K2 Light Curves}

We downloaded K2 light curves from Campaigns 0 through 19, as produced by \citet{VanderburgJohnson2014}, from their \href{https://lweb.cfa.harvard.edu/~avanderb/k2.html}{website}. To match the EPIC IDs of K2 targets with TIC IDs, we used the catalogue match described in \citet{Dotson2020}. Overall, we estimate $\sim$30,000 -- 50,000 K2 targets overlap with each TESS sector across Sectors 42 -- 46, down to the QLP magnitude limit of $T = 13.5$ mag. The K2 light curves were first detrended by a Basis Spline on a per-campaign basis following \citet{VanderburgJohnson2014}, and then normalized and combined through different campaigns. We computed the median absolute median error (MAD) of each campaign to approximate the error of the light curves following the Gaussian assumption. We then iteratively masked out points for each campaign with flux larger than 3-sigma plus the median flux before further processing and plotting.

\subsection{Transit Search}

Before running our Box Least Squares (BLS) transit search, we combine the K2 light curves with normalized TESS light curves using an offset based on the median value of the light curve. The BLS algorithm is applied using weights as implemented in \texttt{VARTOOLS} \citep{HartmanBakos2016}, where the weights are estimated errors from the per-campaign or per-orbit MAD. We select our BLS settings to reflect that the K2 light curves typically have longer baseline compared to the TESS light curves ($\sim80$ days per campaign), and that the two sets of the observations are taken $3 - 6$ years apart. Instead of searching up to a maximum period of half the observation baseline of TESS observations as in our typical transit search, we search to a maximum period of the full observation baseline of TESS observations. For S42, we searched on a uniform period grid from 0.1 -- 28 days using 200,000 bins, and 500 bins in the phase folded grid. The minimum and maximum allowed transit duration to period ratios are 0.004 and 0.08, respectively. These settings will be updated in future sectors as the TESS observing baseline increases for each target.

\subsection{Triage}

Nominal triage of QLP threshold crossing events is performed only for those around stars brighter than $T = 10.5$ mag in order to constrain the number of planet candidates needing manual review to manageable levels. While we continue to follow this procedure for non-K2 targets, we triage events around all stars with K2 overlap down to the $T = 13.5$ mag limit of QLP. 

Starting in Sector 43, we use different signal-to-pink-noise ratio (S/N) cuts to identify the threshold crossing events that will undergo triage. For non-K2 targets, we use our standard S/N $>$ 9. For K2 targets, we require S/N $>$ 9 as calculated from the combination of K2 and TESS light curves, but we also require that the signal is plausibly visible in the TESS data alone by requiring a TESS-only S/N $>$ 5. We did not have this second requirement for Sector 42 triage, but implemented it after reviewing the output of the K2 + TESS search. Because of the different data characteristics of K2 and TESS light curves, we also use two independent neural networks -- one trained with K2 \citep[Astronet-K2;][]{Dattilo:2019}, and one trained with TESS \citep[Astronet-TESSv2; Moldovan et al, in prep,][]{Yu2019} -- to identify transit-like events from the detected threshold crossing events. If one of the neural networks determines that the signal is an astrophysical eclipsing signal, we pass the signal to a manual inspection stage following the standard TOI process \citep{Guerrero2021}. The score thresholds for passing with Astronet-K2 and Astronet-TESSv2 are 0.3 and 0.19, respectively. As part of the triage process, we also perform a preliminary MCMC fit that estimates transit model parameters. This fit includes the K2 light curves with their appropriate errors when available.

\subsection{Vetting Reports}
\begin{figure}
    \centering
    \gridline{\fig{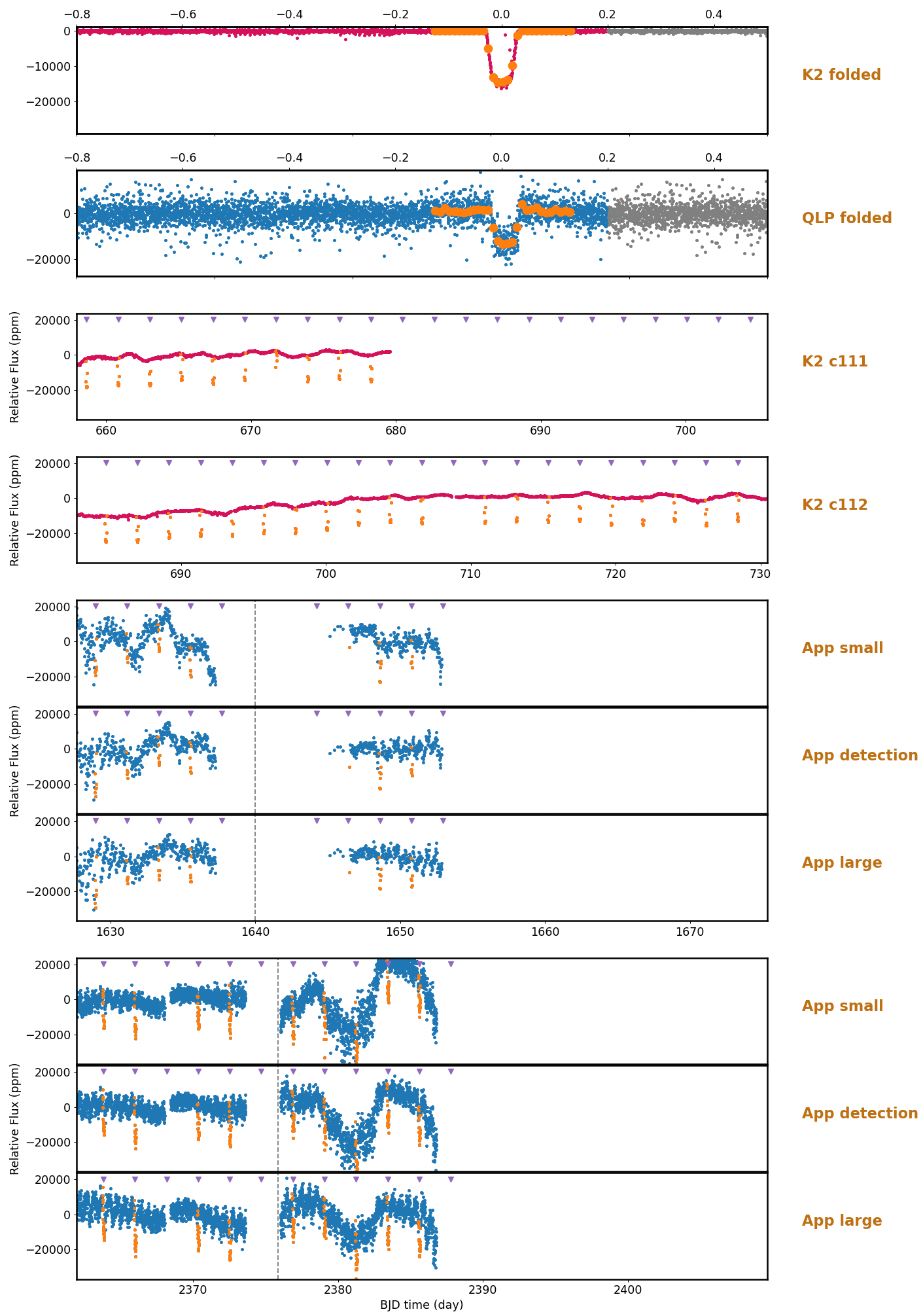}{.6\textwidth}{(a) Top: Folded detrended light curves. Bottom: Raw light curves.}}
    \gridline{\fig{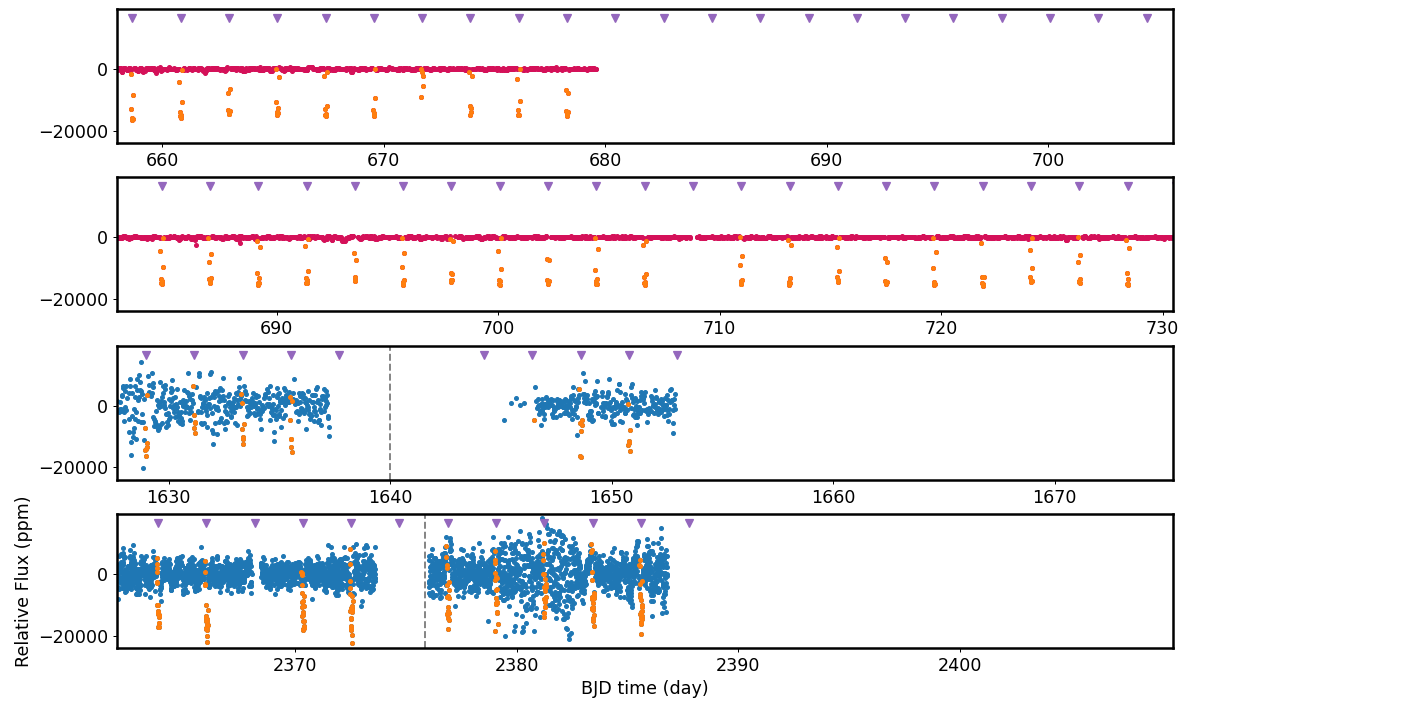}{.6\textwidth}{(b) Detrended light curves.}}
    \caption{Light curves  for TIC 16288184.01 (K2-237b) with the K2 portions in magenta, the QLP portions in blue, and the transits highlighted in orange.}
    \label{fig:report}
\end{figure}

QLP vetting reports are made for all candidates passing manual triage by a QLP operator. The reports for stars that use K2 light curves are slightly changed from the standard report to represent the additional data.
The main changes are the following:
\begin{itemize}
    \item The EPIC ID of the K2 light curve is added in the star section of the summary page. This also includes a note about how the matching EPIC ID was determined (via 2MASS or not). If the match is not made via 2MASS it may be less reliable, so we recommend checking \citet{Dotson2020} for more details.
    \item The light curves plotted in the summary page and MCMC page include both QLP and K2 light curves.
    \item The raw and detrended light curve pages include separate panels with the K2 light curve in similar format with the QLP light curves (see Figure \ref{fig:report}).
\end{itemize}


\section{Summary}

Effective Sector 42 and onward, QLP underwent the following changes to enable analysis of targets observed by both K2 and TESS:

\begin{itemize}
    \item Perform BLS search on combined K2 + QLP data.
    \item Use Astronet-K2 to triage candidates from K2 + QLP.
    \item Reflect K2 data in vetting reports.
\end{itemize}

\section{Acknowledgements}
These data release notes provide processing updates from the Quick Look Pipeline (QLP) at the TESS Science Office (TSO) at MIT. QLP extracts and detrends light curves from TESS Full-Frame Images (FFIs), searches light curves for transits, and produces vetting reports for promising planet candidates. The full QLP procedure is described in \citet{Huang2020}. This work makes use of High-Level Science Products (HLSPs) stored on the Mikulski Archive for Space Telescopes (MAST). The TESS mission is funded by NASA’s Science Mission Directorate.

\bibliography{refs}

\end{document}